\documentclass[12pt]{iopart}

\begin{document}
\def\be{\begin{equation}}

\def\ee{\end{equation}}

\def\bc{\begin{center}} 
\def\ec{\end{center}}
\def\bea{\begin{eqnarray}}

\def\eea{\end{eqnarray}}

\title[Percolation transition in correlated hypergraphs]{Percolation
  transition in correlated hypergraphs}

\author{Serena Bradde}
\address{SISSA, via Beirut 2/4, 34014, Trieste, Italy and INFN, Via Valerio 2, Trieste, Italy}
\ead{bradde@sissa.it}
\author{Ginestra Bianconi}
\address{The Abdus Salam International Center for Theoretical Physics, Strada Costiera 11, 34014, Trieste, Italy}
\ead{gbiancon@ictp.it}
\begin{abstract}
Correlations are known to play a crucial role in determining the structure of
complex networks. Here we study how their presence affects the computation of
the percolation threshold in random hypergraphs.
In order to mimic the correlation in real network, we build
hypergraphs from a generalized hidden variable ensembles and
we study the percolation transition by mapping this
problem to the fully connected Potts model with heterogeneous couplings.
\end{abstract}

\maketitle

\section{Introduction}
In the last decade the topological properties of networks have attracted a large 
interest \cite{Albertbarabasi,Newman_rev,Vito,Vespignani}, mainly driven by the emergence of
novel dynamical effects in random processes defined on them \cite{Dorogovtsev,Alain}.
The impact of this research is wide and has important consequences in 
the domains of biology, socio-economical theories and technological infrastructure
design. It has been shown that complex networks can display a universal behavior
which strongly affects the dynamics of statistical models built on them.

A major example is provided by the percolation transition \cite{Attack,Cohen1,Cohen2,Newman_bicom},
one of the most famous emergent collective phenomena that can 
be defined on complex networks.  
Its dependence on degree distribution, degree-correlations and
directionality of the links has been extensively studied in the last years
\cite{Boguna_per_dir,Doro2,Newman_clust}. 
In particular, the interplay between topological 
features and the nature of the percolation transition has been 
fully investigated within different types of 
random network ensembles \cite{Chung,Kahng_hv,hv1,Boguna_hv,hv2,entropy1,entropy2}.
These constitute null models for networks, each of them being formed 
by graphs sharing with real complex networks a number of structural features, such as degree 
distribution or correlations between neighboring components.
 
Recently, attention has been devoted to the structure of
hypergraphs \cite{hyper1,hyper2}, describing, for example, 
many on-line social and professional communities which collaborate 
in order to give a semantic structure to a set of data.
Among these communities, also named folksonomies, we mention 
Flickr or CiteULike, whose structure is formed by triplets of users, resources and 
tags linked together. 
Such kind of networks show important correlations since the interest of
the user and the subject of the resources (for example a picture for
Flickr and an article for CiteULike) usually have a strong inter-dependence.

These correlations are responsible for the build-up of communities, after projecting
these networks into networks made only by user-user,
resource-resource and tag-tag interactions \cite{hyper2}

In a recent paper \cite{hyper1}, the percolation transition in random
uncorrelated hypergraphs  was characterized providing a first approximation 
to the real hypergraphs properties. In this work we extend 
these results to more general 
ensembles of correlated hypergraphs which can give a better description of 
real social communities. We will show how to build null models 
for hypergraphs based on recent parallel construction introduced 
for  networks \cite{entropy1,entropy2}. 
Moreover, we will derive the percolation threshold of correlated hypergraphs by 
mapping the problem to the solution of a fully connected Potts 
model with heterogeneous couplings \cite{Kahng} following the method 
developed in a recent paper on percolation phase transition in simple
networks \cite{Serena}.

\section{Correlated random hypergraphs}

To mimic the correlation of real hypergraphs we propose to study
randomized ensembles of correlated hypergraphs within the same
theoretical framework developed in \cite{entropy1,entropy2} for
networks.
In these works it has been shown how is possible to correctly define
statistical microcanonical and canonical networks ensembles. 
In particular they show how to obtain the
general relation between $G(N,L)$ random networks ensemble - a collection of 
graphs with fixed number of nodes $N$ and links $L$ - and the $G(N,p)$ one
formed by $N$ nodes linked two by two with probability $p$. In the latter 
case the number of links $L$ fluctuates 
and has a Poissonian distribution with mean $p\times N$. In the thermodynamic 
limit it is known that these two different ensembles share the same statistical
properties when the external parameters, $L$ and $p$, satisfy the following 
relation $p=L/N$.
We call $G(N,L)$ the microcanonical ensemble because it satisfies an hard
constraint on the extensive number of links $L$ - like the energy constraint 
for usual microcanonical ensemble -. While on the other case the $G(n,p)$
ensemble is canonical, in the sense that the Lagrange multiplier $p$ let the
number of links fluctuates fixing only its {\it average} value to $L=p\times N$.
Generalizing this approach to several properties, like i.e. the degree 
sequence or a certain division into different communities, it is possible to 
define complex microcanonical and canonical network ensembles satisfying 
more stringent constraint \cite{entropy1,entropy2}. Here we sketch how this 
general framework can be easily generalized to hypergraphs.

Let us consider a hypergraph  formed  by nodes of different
types $\alpha=1,\ldots, K$ linked in $K$ groups.
For example in Flickr we will have $K=3$ and $\alpha=1,2,3$
indicating respectively agents, pictures and tags.
We assume also, in order to gain in generality, that each node can be
associated to a different feature $r_i^{\alpha}=1,\ldots R^{\alpha}$
indicating a given classification of the nodes. Again, in the 
case of Flickr, the agents can be classified in relation to their
interests or age, the tags in relation to their general meaning and
the pictures in relation of the type of subject is represented.

A given random correlated ensemble of hypergraphs can be defined as
the set of all the hypergraphs which satisfy a number of constraints.

In particular, we choose these constraints to be the number of
hyperlinks each node has and the number of hyperlinks bridging set of
nodes with different features.
Following these prescriptions we can construct microcanonical and canonical
hypergraphs ensembles.

\subsection{Microcanonical hypergraph ensembles}

Let us define a hypergraph by the tensor
$a_{i_1,i_2,\ldots i_K}=1$ if the nodes $(i_1,i_2,\ldots i_K)$ are
linked together and $a_{i_1,i_2,\ldots i_K}=0$ otherwise.
Let's call $N_{\alpha}$ the number of nodes of type $\alpha$.
The networks in the microcanonical hypergraph ensemble will then satisfy the 
following conditions,
\be\label{constraints}
\left\{\begin{array}{rcl} k_{i_{\alpha}}&=&\sum_{\{i_\gamma\}_{\gamma\neq \alpha}}a_{i_1,\ldots
  i_{\alpha},\ldots, i_K}\nonumber \\
A(x_1,x_2,\ldots x_K)&=&\sum_{\{i_\gamma\}}a_{i_1,\ldots
  i_{\alpha},\ldots,
  i_K}\prod_{\gamma}\delta(r_{i_{\gamma}}-x_{\gamma})\end{array}\right.
\ee
with $k_{i\alpha}$ being the hyperdegree of node $i_{\alpha}$ and with
$A(x_1,x_2,\ldots, x_K)$ being the number of hyperedges between nodes
of features $(x_1,x_2\ldots x_K)$.

\subsection{Canonical hypergraph ensembles}

By using the statistical mechanics approach described in
\cite{entropy1,entropy2} it can be shown that networks in the canonical
hypergraph ensembles, satisfying on average the constraints $(\ref{constraints})$,
can be constructed by assigning a hyperlink $a_{i_1,i_2,\ldots, i_K}=1$ with probability
\be
p_{i_1,i_2,\ldots,i_K}=\frac{\theta_{i_1}\theta_{i_2}\ldots
  \theta_{i_K}
  W(r_{i_1},r_{i_2},\ldots,r_{i_K})}{1+\theta_{i_1}\theta_{i_2}\ldots\theta_{i_K}
  W(r_{i_1},r_{i_2},\ldots,r_{i_K})}.
\label{p}
\ee 
If the  constants $\theta_{i_{\alpha}}$ and the tensor
$W(r_1,r_2,\ldots r_K)$ satisfy the following conditions
\be\label{avg}
\left\{ \begin{array}{rcl} \overline{k_{i_{\alpha}}}&=&\sum_{\{i_\gamma\}_{\gamma\neq
    \alpha}}p_{i_1,i_2,\ldots,i_K}\nonumber \\
\overline{A(x_1,x_2,\ldots
  x_K)}&=&\sum_{\{i_\gamma\}}p_{i_1,i_2,\ldots,i_K}\prod_{\gamma=1}^K\delta(r_{i_{\gamma}}-x_{\gamma})\end{array}\right.
\ee
then the  hyperdegrees $k_i$ and the number of hyperedges $A(x_1,x_2,\ldots x_K)$
\bea
\left\{\begin{array}{rcl} k_{i_{\alpha}}&=&\sum_{\{i_\gamma\}_{\gamma\neq \alpha}}a_{i_1,\ldots
  i_{\alpha},\ldots, i_K}\nonumber \\
A(x_1,x_2,\ldots x_K)&=&\sum_{\{i_\gamma\}}a_{i_1,\ldots
  i_{\alpha},\ldots,
  i_K}\prod_{\gamma}\delta(r_{i_{\gamma}}-x_{\gamma})\end{array}\right.
\eea
are Poisson distributed \cite{Boguna_hv} with average $\overline{k_{i_\alpha}}$ and
$\overline{A(x_1,x_2,\ldots x_K)}$ given by $(\ref{avg})$.

In the limit of hypergraphs with a linking probability independent on
the features of the nodes  $r_{i_\gamma}$ 
we obtain that the probability $(\ref{p})$ becomes
\be
p_{i_1,i_2,\ldots,i_K}=\frac{\theta_{i_1}\theta_{i_2}\ldots
  \theta_{i_K} }{1+\theta_{i_1}\theta_{i_2}\ldots\theta_{i_K}}
\label{ens}
\ee 
Moreover, we recover the configuration model for the uncorrelated hypergraphs taking the limit 
$\prod_{\alpha}\theta_{i_{\alpha}}\ll 1 \; \forall\; \{i_{\alpha}\}$. In this case the
hyperedge probability be
\be
p_{i_1,i_2,\ldots,i_K}\simeq {\theta_{i_1}\theta_{i_2}\ldots
  \theta_{i_K} }=
\frac{\overline{k_{i_1}}\ldots \overline{k_{i_K}}}{(\langle{\overline{k}}\rangle N)^{(K-1)}}\,,
\ee
and this last expression describes \emph{uncorrelated} hypergraphs whose
properties have been studied in \cite{hyper1}.
 
\section{Potts Model and percolation transition}

It is a well known result \cite{Fortuin,Wu_rev,Lub} of statistical mechanics that 
the Potts phase transition in a fully connected systems for the number of colors 
$q\rightarrow 1$, describes the bond percolation transition \cite{Kahng,Monasson} 
in Erd\"os R\'enyi $G(N,p)$ random graphs \cite{Bollobas}.
Recently a fully connected  Potts model with heterogeneous couplings \cite{Kahng,Serena}
was introduced in order to study the percolation transition in random
networks with heterogeneous degree distribution and additional structural properties. 
Here we show that such a method can be extended to correlated
canonical hypergraphs.

Instead  of a pure Potts model we consider the following generalized Hamiltonian 
\begin{equation}
 H=-\sum_{i_1 i_2 \ldots i_K} J_{i_1 i_2 \ldots i_K}
 \delta_{s_{i_1}s_{i_2} \ldots s_{i_K}}
\label{H}
\end{equation}
where the summation runs over all the $K$-sets that compose the fully
connected hypergraph and the variables $\{s_{i_\gamma}\}$ are Potts spins taking $q$ values
from $[1,2,\ldots, q]$. We define for later convenience the vector 
$\mathbf{s}_{\alpha}=\{s_{1_{\alpha}},\ldots s_{N_{\alpha}}\}$ where $N_\alpha$ is the number of
sites of a given type $\alpha$. Introducing an external parameter $\beta$ as 
the inverse of temperature, the partition function of the Hamiltonian model 
reads
\begin{equation}
 Z=\sum_{\mathbf{s}_1 \mathbf{s}_2 \ldots \mathbf{s}_K}  e^{-\beta H[\mathbf{s}_{\alpha}]}\,,
\end{equation}
whose cluster expansion gives the following expression
\begin{equation}
 Z=\sum_{H\in K(H)}  \prod_{i_1 i_2 \ldots i_K \in E} v_{i_1 i_2
   \ldots i_K} q^{C(H)}\,.
\label{c}
\end{equation}
The value $C(H)$ is the number of connected components of the hypergraph $H$ and 
$$v_{i_1 i_2\ldots  i_K}=e^{\beta J_{i_1 i_2\ldots i_K}}-1 \simeq \beta J_{i_1,i_2,\ldots i_K}\,,$$
where the last expression is valid in the small $\beta$ limit.
After identifying the coupling
\be
\beta J_{i_1,i_2,\ldots i_K}=\frac{p_{i_1i_2 \ldots
    i_K}}{1-p_{i_1i_2\ldots i_K}}=\theta_{i_1}\theta_{i_2}\ldots
  \theta_{i_K} W(r_{i_1},r_{i_2},\ldots,r_{i_K})
\label{cond}
\ee
then the sum in (\ref{c}) is a weighted sum over random hypergraphs
$H$ in which each link has probability $p_{i_1,i_2,  \ldots,i_K}$ reported in equation (\ref{p}).
If the condition $(\ref{cond})$ holds, the transition of the Hamiltonian
model $(\ref{H})$ for $q\rightarrow 1$ coincides with the percolation transition in
the random correlated hypergraph ensemble.
In the next section we will solve the Potts model providing the
percolation condition for hypergraphs in the ensemble $(\ref{ens})$ 
that could be written in the form 
\be
\det(\Xi)=0
\ee
with the matrix $\Xi$ to be determined in the following.

\subsection{Solving Potts model}

In this subsection we solve the Potts model $(\ref{H})$, in the generic case,
for arbitrary $K$ and $J_{i_1,i_2,\ldots i_K}$ given by $(\ref{cond})$.
We introduce  the order parameter $c^{\alpha}_{\theta,r}(s)$ indicating
the  fraction of nodes of type $\alpha=1,\ldots K$ associated to the
`hidden variable' $\theta$ and the feature $r$, having Potts spin
equal to $s$
\begin{equation}
 c^\alpha_{\theta,r}(s)=\frac{1}{N_\alpha p^{\alpha}(\theta,r)} \sum_{i_\alpha} \delta_{s,s_{i_\alpha}} \delta(\theta,\theta_{i_\alpha})\delta_{{r},{r}_{i_{\alpha}}}.
\end{equation}
where $p^\alpha(\theta,r)$ is the probability distribution to have a node of $\alpha$-type 
with local properties describes by variables $\theta$ and $r$. We have 
introduced for convenience the Dirac delta function, with its 
proper normalization 
$\int d\theta\; \delta(\theta)=1$ and the Kronecker's delta 
$\sum_i\;\delta_{i,0}=1$ 
defined as
\be
\delta(\theta,\theta')=\left\{\begin{array}{ll}
0 & \theta\neq \theta'\\
\infty & \theta=\theta'
\end{array}
\right.
\qquad
\delta_{s,s'}=\left\{\begin{array}{ll}
0 &  s\neq s'\\
1 &  s=s'
\end{array}
\right.
\ee
Noticing the symmetry of the Hamiltonian in terms of $c^\alpha_{\theta,r}(s)$
\begin{equation}
 H[\{c^{\alpha}_{\theta,r}(s)\}]=-\prod_{\gamma}N_{\gamma}\sum_{s,\;
   \{r\}}\prod_{\gamma}\left[\int  d\theta^\gamma  p^\gamma(\theta^\gamma,r^\gamma)c^{\gamma}_{\theta^\gamma, r^\gamma}(s)\right] J(\{\theta\},\{r\})
\end{equation} 
we can express the partition function as a summation over these new variables
\begin{equation}
 Z=\sum_{\{c_{\theta ,r}^\alpha (s)\}} e^{- \beta F[\{c^\alpha_{\theta,r}(s)\}]}
\end{equation}
where the free energy function is given by 
\bea
\beta F[\{c^\alpha_{\theta, r }(s)\}] =&&   \beta H[\{c^{\alpha}_{\theta,r}(s)\}]\nonumber\\
&& +\sum_\alpha N_{\alpha}\int d\theta^{\alpha} \sum_{r^{\alpha}}{p^\alpha(\theta^\alpha,r^\alpha)} c^\alpha_{\theta^\alpha,r^\alpha}(s) \ln c^\alpha_{\theta^\alpha,r^\alpha}(s)
\eea

The phase transition of the Potts model is determined by the point at
which the free energy becomes unstable respect to variation of the
order parameters around the symmetric solution 
\be
c^{\alpha}_{\theta,r}=\frac{1}{q}.
\label{sym}
\ee
In order to determine the stability of the free energy we evaluate the
Hessian ${\cal H}$ of components
\be\label{eq:hessiandef}
{\cal H}^{\{a\},\{a'\}}(s,s')=\frac{\partial \beta F}{\partial c_{\{a\}}(s)\partial c_{\{a'\}}(s')}
\ee
where we indicate by $\{a\} (\{a'\})$ the triplets $\{\alpha,\theta^{\alpha},r^{\alpha}\} (\{\alpha,\theta^{\alpha'},r^{\alpha'}\})$. Making explicitly the calculation, the previous equation 
(\ref{eq:hessiandef}) reads
\bea
\hspace{-15mm}{\cal H}^{\{a\},\{a'\}}(s,s') &=& \delta_{s,s'}\left\{\delta_{\{a\},\{a'\}}
\frac{N_{\alpha} p^\alpha(\theta^{\alpha},r^{\alpha})}{c^\alpha_{\theta^{\alpha},r^{\alpha}}(s)}- p^\alpha(\theta^{\alpha},r^{\alpha})p^{\alpha'}(\theta'^{\alpha'},r'^{\alpha'})\right. \times \nonumber \\
&&\left.\prod_{\gamma}N_{\gamma}  \prod_{\gamma\neq
  \alpha,\alpha'}\left[\int d\theta^{\gamma} \sum_{r^\gamma}p^\gamma(\theta^\gamma,r^{\gamma}) c^\gamma_{\theta^{\gamma},r^{\gamma}}(s) \right]{\beta J(\{\theta\},\{r\})}\right\}\,,
\eea
with the following definition of $\delta_{\{a\},\{a'\}}=\delta_{\alpha,\alpha'}\;\delta(\theta^\alpha,\theta^{\alpha'})\;\delta_{r^\alpha,r^{\alpha'}}$.
Taking $J(\{\theta\},\{q\})$ from equation $(\ref{cond})$, we obtain that the eigenvalue problem
associated to the Hessian matrix is
\bea
\label{eig}
\hspace*{-20mm}\big[\lambda+N_{\alpha}p^\alpha(\theta^{\alpha},r^{\alpha})
q\big] { \bf e}(\{a\})&=&\frac{N_\alpha
  p^\alpha(\theta^{\alpha},r^{\alpha})}{q^{K-2}}\sum_{\alpha'\neq
  \alpha}\left\{
\prod_{\gamma\neq \alpha, \alpha'} \left[\int d\theta^{\gamma}\sum_{r^{\gamma}} N_\gamma
p^\gamma(\theta^{\gamma},r^{\gamma})\right] \right.\times\nonumber \\
&&\left. \int
d\theta^{\alpha'}
\sum_{r^{\alpha'}}p^{\alpha'}(\theta^{\alpha'},r^{\alpha'})\beta J(\{\theta\},\{r\}){\bf e}(\{a'\})\right\}.  
\eea
where $\lambda$ and ${\bf e}(\{a\})$ are respectively the eigenvalue
and the eigenvector of the problem. 
The equation $(\ref{eig})$ can be written as 
 \begin{equation}
{\bf e}(\{a\})= \frac{N_\alpha  p^\alpha(\theta^{\alpha},r^{\alpha})}
{(\lambda+N_{\alpha} p^\alpha(\theta^{\alpha},r^{\alpha}) q)q^{K-2}
}{\bf \Delta}(\{a\})
\label{1}
\end{equation}
with the  ${\bf \Delta}(\{a\})$ defined as 
\bea
 {\bf \Delta}(\{a\})&=&\sum_{\alpha'\neq \alpha}\left\{
\prod_{\gamma\neq\alpha, \alpha'} \left[\int d\theta^{\gamma}\sum_{r^{\gamma}} N_\gamma
p^\gamma(\theta^{\gamma},r^{\gamma})\right]\right.\times \nonumber \\
&&\left. N_{\alpha'}\int
d\theta^{\alpha'}
\sum_{r^{\alpha'}}p^{\alpha'}(\theta^{\alpha'},r^{\alpha'})\beta
J(\{\theta\},\{r\}){\bf e}(\{a'\})\right\}\,.  
\label{2}
\eea
The symmetric solution becomes unstable when the maximal eigenvalue
of the Hessian problem becomes positive. Therefore, in order to
determine the critical point of the Potts model for
$q\rightarrow 1$ we consider
equations $(\ref{1})$ and $(\ref{2})$ when the eigenvalues vanishes
$\lambda=0$.
If we take into account the explicit form of the coupling constant
given by Eq. (\ref{cond}), we find that the vector ${\bf \Delta}$ takes
the form
$\Delta(\{a\})=\theta^{\alpha}{\mathbf{v}}_{\alpha,r^{\alpha}}$ where the
${\bf v}$'s satisfy the linear system of equations 
\be
\Xi {\mathbf{v}}=0
\ee
with the matrix $\Xi_{\{\alpha,r^{\alpha}\}\{\alpha',r^{\alpha'}\}}$ given by  
\bea
\label{eq:matrix}
\hspace*{-20mm}\Xi_{\{\alpha,r^{\alpha}\}\{\alpha', r^{\alpha'}\}}&=&-\delta_{\alpha,\alpha'}\delta_{r^{\alpha},r^{\alpha'}}+[1-\delta_{\alpha,\alpha'}\delta_{r^{\alpha},r^{\alpha'}}] \prod_{\gamma\neq\alpha, \alpha'} \left[\int d\theta^{\gamma}\sum_{r^{\gamma}} N_\gamma p^\gamma(\theta^{\gamma},r^{\gamma})\theta^{\gamma}\right]\nonumber \\
&&\hspace*{-20mm}\times
 \int
d\theta^{\alpha'}N_{\alpha'}p^{\alpha'}(\theta^{\alpha'},r^{\alpha'})(\theta^{\alpha'})^2
W(r_1,\ldots, r^{\alpha},\ldots, r^{\alpha'},\ldots,r_K).  
\eea
 Therefore the condition determining the critical point of the
 Potts model for $q\rightarrow 1$ comes from the vanishing of the determinant, 
i.e. $\det\Xi=0$.

\subsection{Simplified cases}

In the simplified case in which the linking distribution do not depend
on the communities $\{r\}$ we have that 
${\bf v}(\alpha,r^{\alpha})={\bf u}_{\alpha}$.
In fact  the  coupling constant in the Potts model  given by ($\ref{cond}$) becomes simply
\be
J_{i_1,i_2,\ldots i_K}=\theta_{i_1}\theta_{i_2}\ldots \theta_{i_K}
\ee 
then the solution  ${\bf v}(\alpha,r^{\alpha})$ can be expressed as
\be
{\bf v}(\alpha,r^{\alpha})=u^{\alpha}.
\ee
Using the definition  $(\ref{eq:matrix})$  we get a non null solution for the  ${\bf u}$'s if and only if the  condition $\det \Phi=0$ is satisfied, with $\Phi$ defined as
\be
\Phi_{\alpha,\alpha'}=-\delta_{\alpha,\alpha'}+(1-\delta_{\alpha,\alpha'})N_{\alpha'}\langle\theta^2\rangle_{\alpha'}\prod_{\gamma\neq\alpha,\alpha'} \left[N_{\gamma}\langle \theta\rangle_{\gamma}\right].
\ee
In the case $K=3$ this condition reduces to the following relation
\be\label{eq:cond}
2\pi_{12}\pi_{23}\pi_{31} + \pi_{12}\pi_{32}+\pi_{13}\pi_{23}+\pi_{21}\pi_{31}-1=0
\ee
with the proper identification 
\be
\pi_{\alpha,\alpha'}=N_{\alpha}\langle \theta^2\rangle_{\alpha}N_{\alpha'}\langle
\theta\rangle_{\alpha'}. 
\ee
The formula (\ref{eq:cond}) is valid in the general case of 
hypergraphs that show non trivial correlation between nodes. 
We recover as a special case the percolation condition in the 
uncorrelated hypergraphs, just
remembering the relation between the variables $\theta$'s 
and the mean site connectivity 
$\theta_i=\overline{k_i}/(\langle \overline{k} \rangle N)^{2/3}$\,.
Therefore, using the previous condition, the $\pi$'s are given by  
\be
\pi_{\alpha,\alpha'}=\frac{\langle k(k-1)\rangle_{\alpha}}{\langle
  k\rangle }\,
\ee
and after some algebra we obtain the condition for the percolation 
already found in
\cite{hyper1}
\be
\frac{\langle k\rangle_1 }{\langle k^2\rangle_1}+\frac{\langle
  k\rangle_2}{\langle k^2\rangle_2}+\frac{\langle k\rangle_3}{\langle k^2\rangle_3}=2
\ee
\section{Conclusions}

Correlations account for the non-trivial structure of complex networks
and must play a significant role also in the characterization of hypergraphs
describing folksonomies. 
In this paper we have studied ensembles of correlated hypergraphs which
can be used to model the interactions between different types of nodes in 
real complex hypergraphs.
We determined the percolation threshold by mapping this problem
to a fully connected Potts model with heterogeneous couplings.
Our approach extends the present knowledge on percolation in
uncorrelated hypergraph. Future development will link these findings
to the study and characterization of real folksonomies and to the
analysis of the robustness of the giant component phase against the 
removal of nodes or hyperedges.

\section{Acknowledgments}
This paper was supported by the project IST STREP GENNETEC contract No.034952 and
by MIUR grant 2007JHLPEZ.\\

\bibliographystyle{unsrt}
\bibliography{biblio2}

\begin{thebibliography}{10}

\bibitem{Albertbarabasi}
R.~Albert and A.~L. Barab\'asi.
\newblock Statistical mechanics of complex networks.
\newblock {\em Rev. Mod. Phys.}, 74(1):47--97, Jan 2002.

\bibitem{Newman_rev}
M.~E.~J. Newman.
\newblock The structure and function of complex networks.
\newblock {\em SIAM Review}, 45:167, 2003.

\bibitem{Vito}
S.~Boccaletti, V.~Latora, Y.~Moreno, M.~Chavez, and D.-U. Hwang.
\newblock Complex networks: Structure and dynamics.
\newblock {\em Phys. Rep.}, 424(4-5):175 -- 308, 2006.

\bibitem{Vespignani}
R.~Pastor-Satorras and A.~Vespignani.
\newblock {\em Evolution and structure of the Internet: A statistical physics
  approach}.
\newblock Cambridge, University Press, 2001, 1st edition, 2004.

\bibitem{Dorogovtsev}
S.~N. Dorogovtsev, A.~V. Goltsev, and J.~F.~F. Mendes.
\newblock Critical phenomena in complex networks.
\newblock {\em Rev. Mod. Phys.}, 80(4):1275, 2008.

\bibitem{Alain}
M.~Barth\'elemy A.~Barrat and A.~Vespignani.
\newblock {\em Dynamics Processes on Complex Networks}.
\newblock Cambridge, University Press, 2001, 1st edition, 2008.

\bibitem{Attack}
R.~Albert, H.~Jeong, and A.-L. Barab\'asi.
\newblock Error and attack tolerance of complex networks.
\newblock {\em Nature}, 406:378, 2000.

\bibitem{Cohen1}
R.~Cohen, K.~Erez, D.~ben Avraham, and S.~Havlin.
\newblock Resilience of the internet to random breakdowns.
\newblock {\em Phys. Rev. Lett.}, 85(21):4626--4628, Nov 2000.

\bibitem{Cohen2}
R.~Cohen, K.~Erez, D.~ben Avraham, and S.~Havlin.
\newblock Breakdown of the internet under intentional attack.
\newblock {\em Phys. Rev. Lett.}, 86(16):3682--3685, Apr 2001.

\bibitem{Newman_bicom}
M.~E.~J. Newman and G.~Ghoshal.
\newblock Bicomponents and the robustness of networks to failure.
\newblock {\em Phys. Rev. Lett.}, 100:138701, 2008.

\bibitem{Boguna_per_dir}
M.~Bogu\~n\'a and M.~A. Serrano.
\newblock Generalized percolation in random directed networks.
\newblock {\em Phys. Rev. E}, 72:016106, 2005.

\bibitem{Doro2}
A.~V. Goltsev, S.~N. Dorogovtsev, and J.~F.~F. Mendes.
\newblock Percolation on correlated networks.
\newblock {\em Phys. Rev. E}, 78(5):051105, 2008.

\bibitem{Newman_clust}
M.~E.~J. Newman.
\newblock Random graphs with clustering.
\newblock {arXiv.org:0903.4009}, 2009.

\bibitem{Chung}
F.~Chung and L.~Lu.
\newblock Connected components in random graphs with given expected degree
  sequences.
\newblock {\em Annals of Combinatorics}, 6(2):125, 2002.

\bibitem{Kahng_hv}
B.~Kahng K.~I.~Goh and D.~Kim.
\newblock Universal behavior of load distribution in scale-free networks.
\newblock {\em Phys. Rev. Lett.}, 87:278701, 2001.

\bibitem{hv1}
G.~Caldarelli, A.~Capocci, P.~De~Los~Rios, and M.~A. Mu\~noz.
\newblock Scale-free networks from varying vertex intrinsic fitness.
\newblock {\em Phys. Rev. Lett.}, 89(25):258702, Dec 2002.

\bibitem{Boguna_hv}
M.~Bogu\ n\'a and R.~Pastor-Satorras.
\newblock Class of correlated random networks with hidden variables.
\newblock {\em Phys. Rev. E}, 68:036112, 2003.

\bibitem{hv2}
J.~Park and M.~E.~J. Newman.
\newblock Statistical mechanics of networks.
\newblock {\em Phys. Rev. E}, 70:066117, 2004.

\bibitem{entropy1}
G.~Bianconi.
\newblock Entropy of randomized network ensembles.
\newblock {\em Europhys. Lett.}, 81:28005, 2008.

\bibitem{entropy2}
G.~Bianconi.
\newblock Entropy of network ensembles.
\newblock {\em Phys. Rev. E}, 79(3):036114, 2009.

\bibitem{hyper1}
G.~Ghoshal, V.~Zlati\'c, G.~Caldarelli, and M.~E.~J. Newman.
\newblock Random hypergraphs and their applications.
\newblock {arXiv.org:0903.0419}, 2009.

\bibitem{hyper2}
V.~Zlati\'c, G.~Ghoshal, and G.~Caldarelli.
\newblock Hypergraph topological quantities for tagged social networks.
\newblock {arXiv.org:0905.0976}, 2009.

\bibitem{Kahng}
B.~Kahng D.~S.~Lee, K. I.~Goh and D.~Kim.
\newblock Evolution of scale-free random graphs: Potts model formulation.
\newblock {\em Nucl. Phys. B}, 696(3):351, 2004.

\bibitem{Serena}
S.~Bradde and G.~Bianconi.
\newblock Percolation transition and distribution of connected components in
  generalized random network ensembles.
\newblock {\em J. Phys. A: Math. Theor.}, 42(19):195007, 2009.

\bibitem{Fortuin}
C.~M. Fortuin and P.~W. Kasteleyn.
\newblock On the random-cluster model : I. introduction and relation to other
  models.
\newblock {\em Physica}, 57(4):536 -- 564, 1972.

\bibitem{Wu_rev}
F.~Y. Wu.
\newblock The potts model.
\newblock {\em Rev. Mod. Phys.}, 54(1):235--268, Jan 1982.

\bibitem{Lub}
T.~C. Lubensky.
\newblock {\em Thermal and geometrical critical phenomena in random systems}.
\newblock North-Holland Pub. Co. ; sole distributors for the USA and Canada,
  Elsevier North-Holland, Amsterdam ; New York : New York :, 1979.

\bibitem{Monasson}
A.~Engel, R.~Monasson, and A.~K. Hartmann.
\newblock On large deviation properties of erdos-renyi random graphs.
\newblock {\em J. Stat. Phys.}, 117:387, 2004.

\bibitem{Bollobas}
B.~B. Bollobas.
\newblock {\em Random graphs}.
\newblock Cambridge, University Press, 2001, 2nd edition, 1985.

\end{thebibliography}

\end{document}